\pdfoutput=1 
\documentclass[%
 reprint,
 amsmath,amssymb,
 aps,
 prl,
 superscriptaddress,
]{revtex4-1}

\usepackage{graphicx}% Include figure files
\usepackage{dcolumn}% Align table columns on decimal point
\usepackage{bm}% bold math
\usepackage{siunitx}% si unit
\usepackage[breaklinks=true,pdfborder={0 0 0},pdftex]{hyperref}% add hypertext capabilities

\usepackage{lineno}

\newcommand{\Sk}[1]{\mathrm{Sk}[#1]}  %skewness
\newcommand{\Ku}[1]{\mathrm{Ku}[#1]}  %skewness

\begin{document}

\title{Kardar-Parisi-Zhang Interfaces with Inward Growth }% 
\author{Yohsuke T. Fukai}
\email{ysk@yfukai.net}
\affiliation{%
	Department of Physics, the University of Tokyo
}%
\affiliation{%
	Department of Physics, Tokyo Institute of Technology
}%
\author{Kazumasa A. Takeuchi}%
\email{kat@kaztake.org}
\affiliation{%
	Department of Physics, Tokyo Institute of Technology
}%

\date{\today}

\begin{abstract}
We study the $(1+1)$-dimensional Kardar-Parisi-Zhang (KPZ) interfaces growing inward from ring-shaped initial conditions, experimentally and numerically, using growth of a turbulent state in liquid-crystal electroconvection and an off-lattice Eden model, respectively. To realize the ring initial condition experimentally, we introduce a holography-based technique that allows us to design the initial condition arbitrarily. Then, we find that fluctuation properties of ingrowing circular interfaces are distinct from those for the curved or circular KPZ subclass and, instead, are characterized by the flat subclass. More precisely, we find an asymptotic approach to the Tracy-Widom distribution for the Gaussian orthogonal ensemble and the $\text{Airy}_1$ spatial correlation, as long as time is much shorter than the characteristic time determined by the initial curvature. Near this characteristic time, deviation from the flat KPZ subclass is found, which can be explained in terms of the correlation length and the circumference. Our results indicate that the sign of the initial curvature has a crucial role in determining the universal distribution and correlation functions of the KPZ class.
\end{abstract}

\maketitle

%%%%%%%%Introduction%%%%%%%%

The concept of universality class,
 which was originally introduced for equilibrium critical phenomena,
 is now extending its applicability to scale-invariant phenomena
 in out-of-equilibrium systems.
It is then fundamental to ask whether one can characterize
 an out-of-equilibrium universality class as deeply as
 an equilibrium counterpart -- such as the celebrated Ising class --
 and, if yes, whether it has something conceptually new in it.
In this context, recent developments on the Kardar-Parisi-Zhang (KPZ) class
 \cite{kardar_dynamic_1986, barabasi_fractal_1995,[{For recent reviews on theoretical studies, see }][]kriecherbauer_pedestrians_2010,*corwin_kardarparisizhang_2012,*quastel_one-dimensional_2015,*halpin-healy_kpz_2015,*[][ etc.]sasamoto_1d_2016},
 which describes fluctuations of growing interfaces as a notable example  are particularly relevant.
For the one-dimensional case, the KPZ class
 became analytically tractable by exact solutions to simple integrable models
 \cite{kriecherbauer_pedestrians_2010},
 which moreover showed intriguing \textit{geometry dependence}
 as we shall state below.

When a $d$-dimensional interface separates
 two distinct regions in $(d+1)$-dimensional space,
 and one region expands into the other  in the presence of noise, 
 the interface typically develops scale-invariant fluctuations,
 without fine tuning of the system parameters \cite{barabasi_fractal_1995}.
The KPZ class describes such an interface under generic conditions,
 without long-range interactions and conservation laws.
It is represented by the KPZ equation \cite{kardar_dynamic_1986},
 which reads, for interface height $h\left({x,t}\right)$
 at lateral position $x\in\mathbb{R}^d$ and time $t\in\mathbb{R}$,
\begin{equation}
\partial_t h(x,t) = \nu\nabla^2{h(x,t)}+{\frac{\lambda}{2}}\left({\nabla h(x,t)}\right)^2+\eta(x,t).
\end{equation}
Here, $\eta$ is a Gaussian noise satisfying $\left<\eta(x,t)\right> = 0$ and $\left<\eta(x,t)\eta(x',t')\right>=D\delta(x-x')\delta(t-t')$ and $\left<\cdots\right>$ denotes the ensemble average.
Then the height $h(x,t)$ develops nontrivial fluctuations
 with amplitude $\sim t^\beta$ and correlation length $\sim t^{1/z}$.
These exponents are universal
 -- thus, the KPZ class is defined --
 as demonstrated by numerous growth models \cite{barabasi_fractal_1995,kriecherbauer_pedestrians_2010} and
 a growing variety of experiments
 \cite{[See ][ for a review on experiments.]takeuchi_experimental_2014}:
 colony growth of living cells \cite{wakita_self-affinity_1997,huergo_morphology_2010,*huergo_dynamics_2011,*huergo_growth_2012},
 paper combustion \cite{maunuksela_kinetic_1997,*myllys_kinetic_2001}, 
 liquid-crystal electroconvection \cite{takeuchi_universal_2010,takeuchi_growing_2011,takeuchi_evidence_2012},
 particle deposition
 underlying coffee ring effect \cite{yunker_CLE_2013},
 chemical wave fronts \cite{atis_experimental_2015}, etc.
Theoretically, KPZ also arises in directed polymer problems \cite{barabasi_fractal_1995,kriecherbauer_pedestrians_2010},
 fluctuating hydrodynamics \cite{spohn_fluctuating_2016},
 incompressible active matter \cite{chen_mapping_2016},
 quantum entanglement under random unitary dynamics \cite{nahum_quantum_2017},
 and so on.

For the one-dimensional KPZ class, $\beta=1/3$ and $z=3/2$ \cite{forster_large-distance_1977,kardar_dynamic_1986,barabasi_fractal_1995},
so that $h(x,t)$ can be expressed as follows:
\begin{equation}
	h\left( x,t \right) \simeq v_\infty t + \left( \Gamma t \right)^{\frac{1}{3}} \chi\left( x', t \right) 
\label{eq:h_asymptotic}
\end{equation}
where statistical variable $\chi\left({x', t}\right)$ denotes rescaled height, $x'$ is defined by $x':=x/\xi\left( t\right)$ with correlation length $\xi\left(t\right):=\frac{2}{A}\left(\Gamma{t}\right)^{2/3}$, and $v_\infty, A>0$ and $\Gamma$ are nonuniversal parameters.
Recent studies on integrable models then unveiled
 exact statistical properties of $\chi\left({x',t}\right)$ in $t\to\infty$
 \cite{kriecherbauer_pedestrians_2010},
 in particular the distribution and spatial correlation functions,
 which were moreover shown to depend on the global geometry of interfaces.
For example, (1) (globally) flat interfaces growing from a straight line,
 and (2) circular (or curved) interfaces starting from a point (or equivalent)
 were found to show different distribution functions: what arises is
 the largest-eigenvalue distribution of random matrices
 in the Gaussian orthogonal (unitary) ensemble
 [GOE (GUE) Tracy-Widom distribution \cite{anderson_introduction_2010}]
 in the flat (circular) case, respectively \cite{johansson_shape_2000,prahofer_universal_2000,sasamoto_spatial_2005,ferrari_determinantal_2005,sasamoto_one-dimensional_2010,*amir_probability_2011,*calabrese_free-energy_2010,*dotsenko_bethe_2010,calabrese_exact_2011}.
In other words, $\chi$ in Eq.~\eqref{eq:h_asymptotic}
 turned out to converge (in distribution) to different random variables,
 denoted by $\chi_1$ (flat) and $\chi_2$ (circular)
 \cite{prahofer_universal_2000}.
Similarly, the spatial correlation was also shown to be different,
 namely the $\text{Airy}_1$ (flat) \cite{sasamoto_spatial_2005}
 and $\text{Airy}_2$ (circular) \cite{prahofer_scale_2002}
 correlation function.

This geometry dependence turned out to be relevant in real experiments also.
Using liquid-crystal (LC) electroconvection,
 the authors of Refs.\cite{takeuchi_universal_2010,takeuchi_growing_2011,takeuchi_evidence_2012} studied
 growing clusters of turbulence, 
 or more precisely, spatiotemporal chaos
 called the dynamic scattering mode 2 (DSM2),
 which invades another turbulent state, DSM1.
The initial DSM2 cluster was nucleated by ultraviolet (UV) laser pulses
 focused on a point or a line,
 generating a circular or flat interface, respectively. 
These interfaces were indeed shown to exhibit distinct properties,
 as described above for integrable models
 \cite{takeuchi_growing_2011,takeuchi_evidence_2012}.
Therefore, these properties are universal, yet geometry dependent;
 in other words, the KPZ class consists of
 a few different \textit{universality subclasses},
 characterized by the same exponents
 but different distribution and correlation properties.
Importantly, the circular and flat subclasses may even have
 \textit{qualitative} differences, such as algebraic (circular) vs superexponential (flat) decay of the spatial correlation function
 \cite{bornemann_airy1_2008},
 persistence of time correlation in the circular case \cite{takeuchi_evidence_2012,takeuchi_characteristic_2016,	de_nardis_memory_2017}, etc.
Theoretically, another subclass for stationary interfaces was also established
 \cite{kriecherbauer_pedestrians_2010}.
 
\begin{figure}
\includegraphics[width=7.0cm]{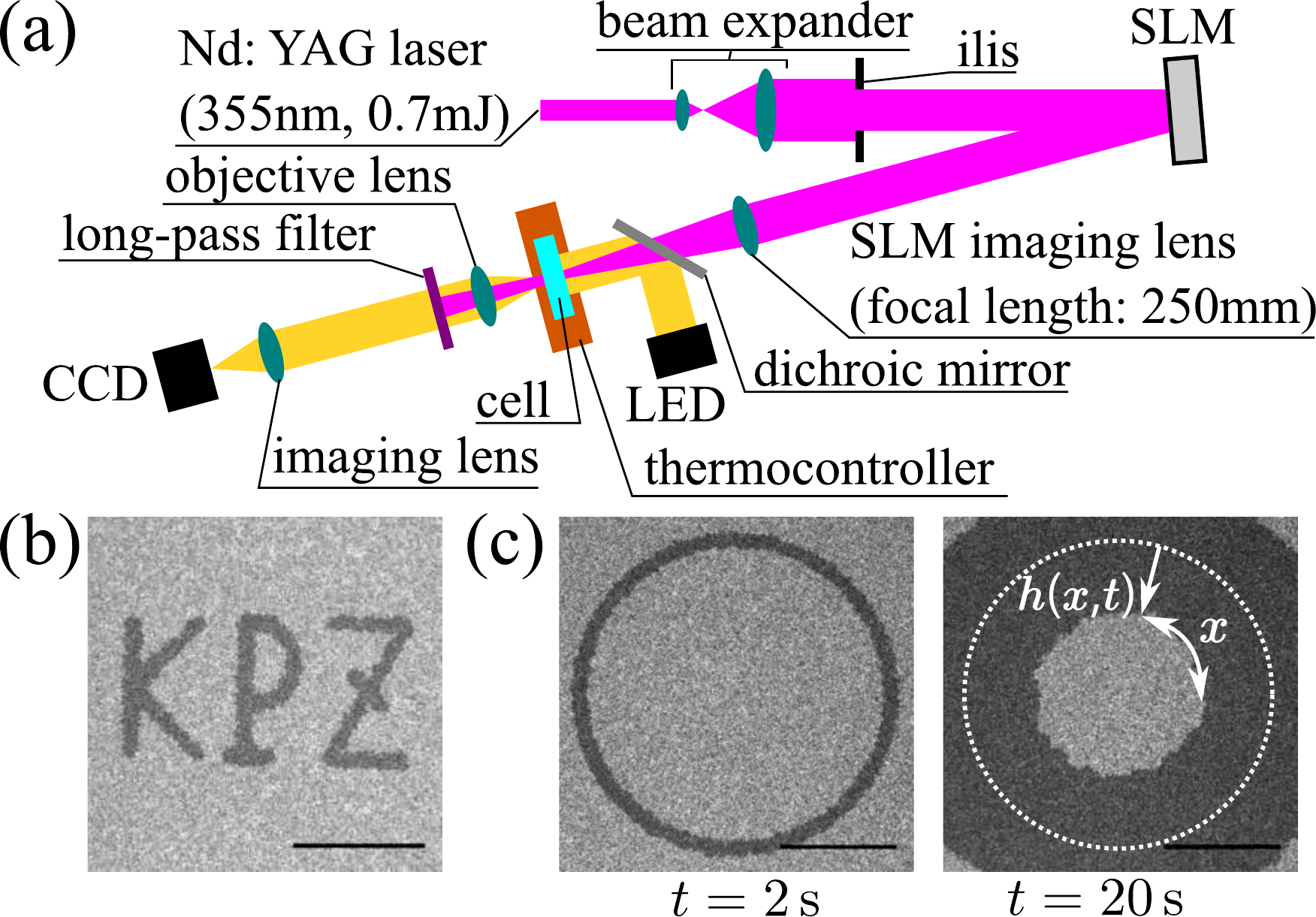}
\caption{\label{fig:schematic} (color online). 
(a) Schematic of the optical setup. Nd:YAG: neodymium-doped yttrium aluminum garnet, SLM: spatial light modulator, CCD: charge-coupled device camera, LED: light-emitting diode. The whole setup is placed in an isothermal chamber as in Ref.\cite{takeuchi_evidence_2012}. (b,c) Images of DSM2, growing from a ``KPZ'' initial condition (b) and a ring with $R_0=\SI{1342}{\um}$ (c). For (c), the elapsed time after laser shooting is indicated below each image. The scale bar is $\SI{1}{\mm}$. The dotted line in the right figure indicates the estimated initial condition. See also Movies S1 and S2 [32].}
\end{figure}

However, in contrast to the detailed knowledge on each subclass,
 relevant parameters that determine the subclass still remain unclear. 
To shed light on this problem experimentally, here we developed
 a holography-based technique to generate
 an \textit{arbitrary} initial condition of the DSM2 interface.
Using this technique, we realized ring-shaped initial conditions
 and measured interfaces growing \textit{inward} from the ring
 (Fig.\ref{fig:schematic}(b)).
Although it is an experimentally natural geometry
 (e.g., coffee ring effect \cite{yunker_CLE_2013}),
 to our knowledge it has never been studied theoretically.
One may guess that this would also lead to the circular subclass,
 because of the presence of the curvature, but it is not clear
 how the \emph{sign} of the initial curvature intervenes.
Our aim is to give a clear answer to this problem.
Numerical simulations were also carried out to corroborate
 the experimental results.

%%%%%%%%Method (LC experiment)%%%%%%%%

The basic experimental setup for the LC electroconvection
 was similar to that used in 
 \cite{takeuchi_universal_2010,takeuchi_growing_2011,takeuchi_evidence_2012}.
We prepared a LC cell,
 which consisted of two glass plates with transparent electrodes,
 separated by spacers of $\SI{12}{\um}$ thickness.
Our LC sample, \textit{N}-(4-Methoxybenzylidene)-4-butylaniline
 doped with $\SI{0.01}{wt.\%}$ of tetra-\textit{n}-butylammonium bromide,
 was introduced to a $\SI{1.5}{\cm}\times\SI{1.5}{\cm}$ area
 enclosed by the spacers. 
The electrodes were coated by \textit{N},\textit{N}-dimethyl-\textit{N}-octadecyl-3-aminopropyltrimethoxysilyl chloride
 to obtain the homeotropic alignment. 
The cutoff frequency separating the conductive and dielectric regimes
 \cite{gennes_physics_1995} was $\SI{1.77(11)}{\kHz}$.
Then, DSM2 can be generated by shooting UV laser pulses to the cell,
 subjected beforehand to a relatively high voltage
 that drives the convection.

In order to realize arbitrary initial conditions,
 one needs to design the beam profile of the UV laser.
To this end, we constructed an optical setup (Fig.\ref{fig:schematic}(a)) using
 a spatial light modulator (SLM, Hamamatsu Photonics, LCOS-SLM X10468-05),
 which allowed us to control the phase profile of the beam
 at the resolution of $792\times{600}$ pixels.
We tuned the phase modulation
 by the iterative Fourier transform algorithm \cite{wyrowski_IFTA_1988}
 until the Fourier-transformed image
 generated by the lens
 had an intensity profile close to the desired one
\footnote{
We noticed that a tiny fraction of light was reflected by SLM
 without phase modulation, and caused uncontrolled DSM2 nucleation.
To avoid this problem, we shifted the effective focal distance
 of the imaging lens,
 by superimposing a phase difference equivalent to a convex lens at SLM.
}.
Thereby, we were indeed able to design the initial DSM2 region
arbitrarily (Fig.\ref{fig:schematic}(b)).

In this Letter, we focus on the case of ring-shaped initial conditions,
 as exemplified in Fig.\ref{fig:schematic}(c).
During the experiment, the cell was attached to a temperature controller,
 whose temperature was maintained at $\SI{25}{\celsius}$
 with fluctuations of about $\pm\SI{0.01}{\celsius}$.
The cell was illuminated by a light-emitting diode.
For each realization,
 we started to apply $\SI{500}{\Hz}$ $\SI{31}{\V}$ AC voltage to the cell,
 shot UV laser pulses to generate a ring-shaped DSM2 region of radius $R_0$,
 and recorded the ingrowing interface through the transmitted light
 by a charge-coupled device camera.
We chose $R_0=\SI{1342}{\um}$, $\SI{1241}{\um}$ and $\SI{826}{\um}$ and
 obtained more than 1500 samples for each case (see Table~SI \cite{supplemental}). 
To define the height $h(x,t)$, we first determined the center of the ring,
 using the ensemble average of the images
 taken at the first frame used in the analysis. 
Then $h(x,t)$ was defined as the radial displacement from the initial ring
estimated from the images at the first frame used in the analysis,
 and $x$ is measured along the circumference,
 whose radius is equal to the mean radius of the interfaces at time $t$
 (Fig.\ref{fig:schematic}(c)). 
The overhangs were averaged out.
%For comparison,
We also obtained flat interfaces from the line initial condition.

%%%%%%%%Method (Eden model)%%%%%%%%

To better understand the experimental results and check the universality,
 we also conducted simulations of ingrowing interfaces
 in an off-lattice Eden model \cite{takeuchi_eden_2012},
 which is an isotropic model known to be in the KPZ class,
 but adapted here to the inward growth.
The model consists of a cluster of round particles with unit diameter,
 which are added one by one stochastically in two-dimensional space.
Starting from an initial condition
 composed of $N$ circularly arranged particles,
 the model evolves following the rules described in \cite{supplemental}.
We simulated ingrowing interfaces with
 $N=8000,16000,32000,100000$ and obtained $2400,3200,3200,1600$
 realizations, respectively.
The initial radius $R_0$ is given by $R_0=N/2\pi$.
We also obtained $3200$ flat interfaces
 using the line initial condition of length $25000$.

%%%%%%%%Result (skewness kurtosis)%%%%%%%%

\begin{figure}
\includegraphics[width=\linewidth]{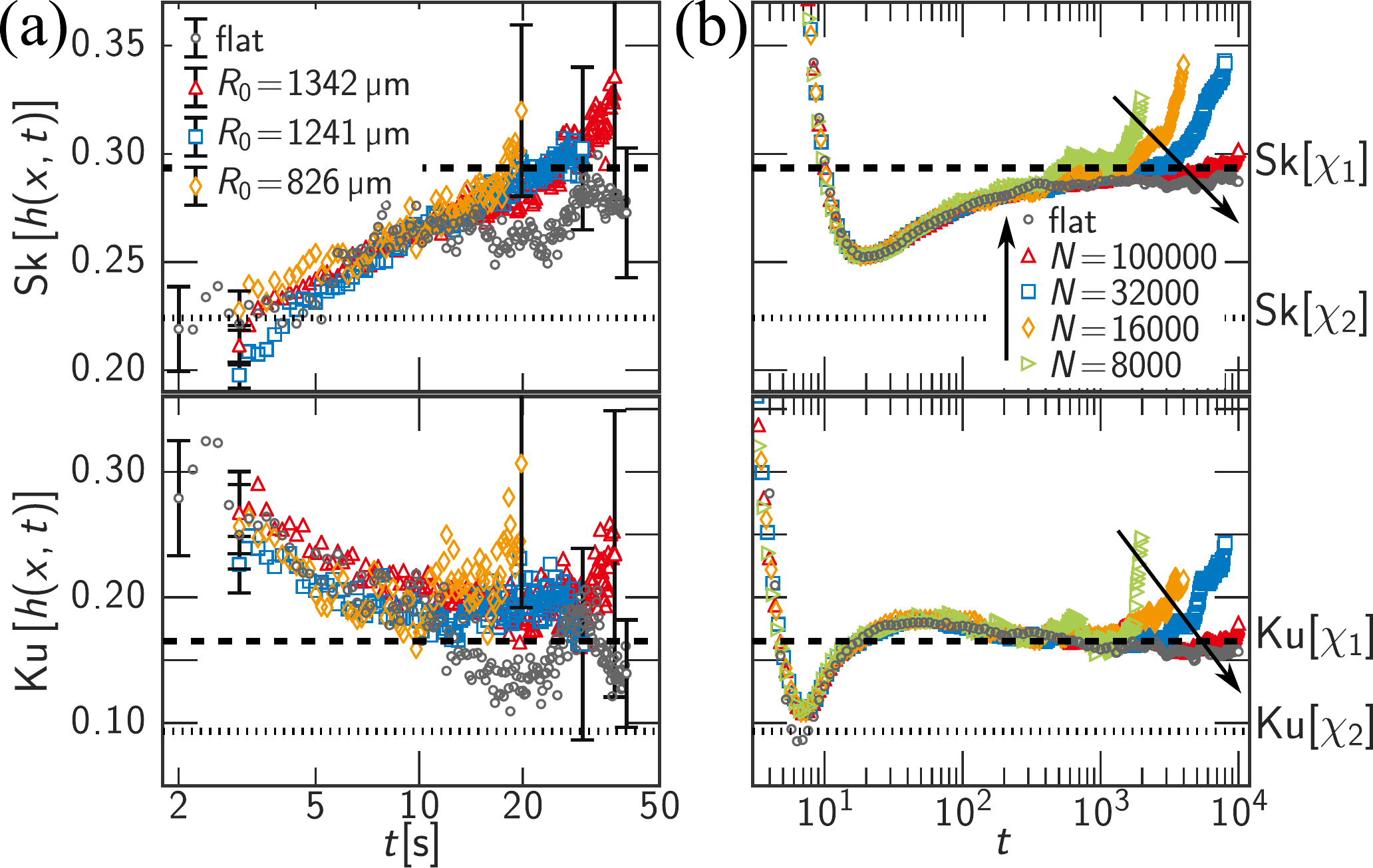}
\caption{\label{fig:skew_kurt} (color online).
Skewness and kurtosis. The values for $\chi_1$ and $\chi_2$
 are shown by the dashed and dotted lines, respectively.
(a) Experimental results. 
Statistical errors are indicated by the error bars,
 shown at the first and last data points.
(b) Numerical results. 
The arrows indicate decreasing initial curvature.
}
\end{figure}

First, we measured the $n$th-order cumulants of $h(x,t)$
\footnote{
When we calculated the cumulants from the experimental data, we evaluated the cumulants of the fluctuations at each $x$ and averaged them, in order to reduce the effect of heterogeneity of the cell and the initial condition.
},
 denoted by $\left<h^n\right>_c$, and evaluated the skewness
 $\Sk{h}:=\left<h^3\right>_c/\left<h^2\right>_c^{3/2}$
 and the kurtosis $\Ku{h}:=\left<h^4\right>_c/\left<h^2\right>_c^2$.
Figure \ref{fig:skew_kurt}(a) and (b) show the results for the LC experiment and the Eden model, respectively.
In both cases, after initial transient, the values for the flat interfaces
 agree with those for $\chi_1$, i.e., the GOE Tracy-Widom distribution, as expected.
Curiously, for the ingrowing cases, too,
 no sign of the circular subclass was found, but the initial transient
 data agreed with those of the flat interfaces. 
For the Eden model (Fig.~\ref{fig:skew_kurt}(b)),
 the skewness and kurtosis then reached and stayed
 for a while at the $\chi_1$ values, until they eventually deviated 
in the direction opposite to the $\chi_2$ values,
at some characteristic times dependent on the initial curvature.
 
%%%%%%%%Result (Non-universal parameters)%%%%%%%%

In order to investigate the cumulants of the rescaled height $\chi\left(x',t\right)$ and its spatial correaltion, we estimated the nonuniversal parameters $v_\infty$ and $\Gamma$ .
Experimentally, we determined the parameters for each set of experiments,
 because of the possible change of the parameter values \cite{takeuchi_evidence_2012}.
For the flat interfaces, we evaluated $v_\infty$
 from the mean growth speed by using $\left<\partial_t{h}\right>\simeq{v_\infty+\mathrm{const.}\times{t^{-2/3}}}$,
 and $\Gamma$ from the variance by fitting 
$\left<h(x,t)^2\right>_ct^{-2/3}\simeq\Gamma^{2/3}\left<\chi_1^2\right>_c+\mathrm{const.}\times{t^{-2/3}}$ \cite{takeuchi_evidence_2012} (Fig.~S1 \cite{supplemental}). 
For the ingrowing interfaces, 
 on the basis of the observation that the early-time behavior of the skewness and kurtosis overlaps with that of the flat interfaces
 (Fig.\ref{fig:skew_kurt}),
 we assumed that time series of individual cumulants, when properly rescaled, also overlap with the flat data,
 and thereby determined $v_\infty$ and $\Gamma$
 (see \cite{supplemental} for details).
 The estimated parameter values are summarized in Table~SI \cite{supplemental}.
For the Eden model,
 since the nonuniversal parameters are expected not to depend on the geometry
 \cite{prahofer_universal_2000,takeuchi_evidence_2012},
 we used the estimates from extensive simulations
 in Ref.\cite{alves_non-universal_2013},
 while using the values in Ref.\cite{takeuchi_eden_2012}
 did not change the conclusion drawn below. For both cases, the parameter $A$ is given by $A=\sqrt{2\Gamma/v_\infty}$
that holds for isotropic systems \cite{takeuchi_evidence_2012}.

%%%%%%%%Result (rescaled quantities)%%%%%%%%

Using these estimates, we obtained the rescaled height
\begin{equation}
 q\left(x',t\right) :=
\left(\Gamma t\right)^{-1/3}\left(h\left(x,t\right) -v_\infty t \right)
 \simeq \chi(x', t)
\end{equation}
 and evaluated the cumulants $\left<q^n\right>_c\simeq\left<\chi(x',t)^n\right>_c$ up to $n=4$.
As for the mean $\left<q\right>$, since it is sensitive to the estimation of $R_0$ and has a large finite-time effect \cite{takeuchi_growing_2011,takeuchi_evidence_2012,takeuchi_eden_2012}, we used instead the mean value of the rescaled velocity
\begin{equation}
 \left< p \left(x',t\right)\right> := \left< \frac{3 t^{2/3}}{\Gamma^{1/3}}\left(\partial_t h\left(x,t\right) -v_\infty \right)\right>
\simeq \left<\chi(x', t)\right>
\end{equation}
to compare with $\left<\chi(x',t)\right>$.

\begin{figure}
\includegraphics[width=\linewidth]{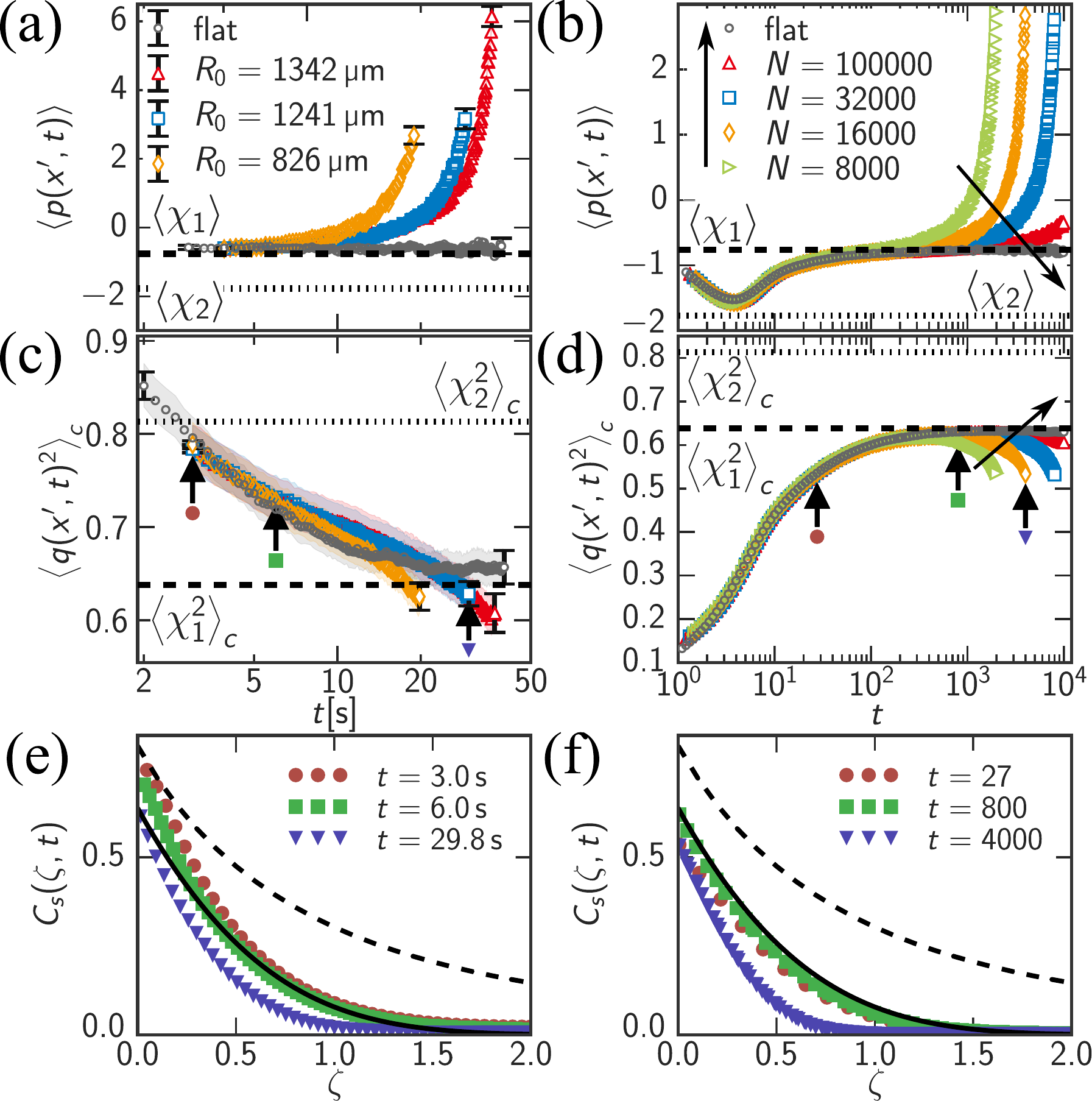}
\caption{\label{fig:rescaled} (color online). 
Rescaled fluctuation properties.
(a-d) Mean rescaled velocity $\left<p(x,t)\right>$ (a,b) and  the variance $\left<q(x,t)^2\right>_c$ (c,d) for the LC experiment (a,c) and the Eden model (b,d). 
The statistical errors are shown by the error bars at the first and last data points in (a,c). The confidence interval associated with the estimation of the nonuniversal parameters is indicated by the shaded area in (c), while it is smaller than the marker size in (a). The arrows in (b,d) indicate decreasing initial curvature. 
The horizontal lines indicate the mean or variance of $\chi_1$ (dashed lines) and $\chi_2$ (dotted lines).
(e,f) The spatial correlation function $C_s\left(\zeta,t\right)$
 for the LC experiment with $R_0=\SI{1241}{\um}$ (e), and for the Eden model with $N=16000$ (f). The solid and dashed lines indicate
 the $\text{Airy}_1$ and $\text{Airy}_2$ correlation function, respectively. 
}
\end{figure}

The experimental and numerical data for 
the mean rescaled velocity $\left<p\right>$ and the variance $\left<q^2\right>_c$ 
are plotted in Figs. \ref{fig:rescaled}(a)-\ref{fig:rescaled}(d).
As expected, the values for the flat interfaces converge to $\left<\chi_1\right>$ and $\left<\chi_1^2\right>_c$, respectively. 
The ingrowing data were found to approach
 the corresponding cumulants of $\chi_1$ in a way similar to the flat case,
 but eventually deviate at some characteristic time,
 which becomes larger for larger $R_0$ (i.e. smaller initial curvature).  In all cases, no sign of approach to $\left<\chi_2\right>$ and $\left<\chi_2^2\right>_c$ was found, similarly to
 the results of the skewness and kurtosis (Fig.\ref{fig:skew_kurt}). Consistent behaviors were found in the third- and fourth-order cumulants (Fig.~S2 \cite{supplemental}), despite relatively large statistical error.
We also measured the two-point spatial correlation function
 $C_s\left(\zeta,t\right):=\mathrm{Cov}\left[q\left(x'+\zeta,t\right),q\left(x',t\right)\right]$ 
 for the ingrowing interfaces [Figs. \ref{fig:skew_kurt}(e) and \ref{fig:skew_kurt}(f)]. 
While $\left<p\right>$ and $\left<q^2\right>_c$ are approaching
 $\left<\chi_1\right>$ and $\left<\chi_1^2\right>_c$, respectively, 
$C_s\left(\zeta,t\right)$ approaches the $\text{Airy}_1$ correlation
 [see the circular and square bullets in Figs. \ref{fig:rescaled}(e) and \ref{fig:rescaled} (f)
 and compare with the black solid line],
 which is the hallmark of the flat subclass. 
After $\left<p\right>$ and $\left<q^2\right>_c$ deviate
 from the values of the flat interfaces,
 $C_s\left(\zeta,t\right)$ also deviates from the $\text{Airy}_1$ correlation
 [the triangle bullets in Figs. \ref{fig:rescaled}(e) and \ref{fig:rescaled}(f)]. 
In any case, $C_s\left(\zeta,t\right)$ was clearly different from the $\text{Airy}_2$ correlation (black dashed line)  for the circular subclass.

%%%%%%%%Discussion (Rescaling time)%%%%%%%%

\begin{figure}
\includegraphics[width=\linewidth]{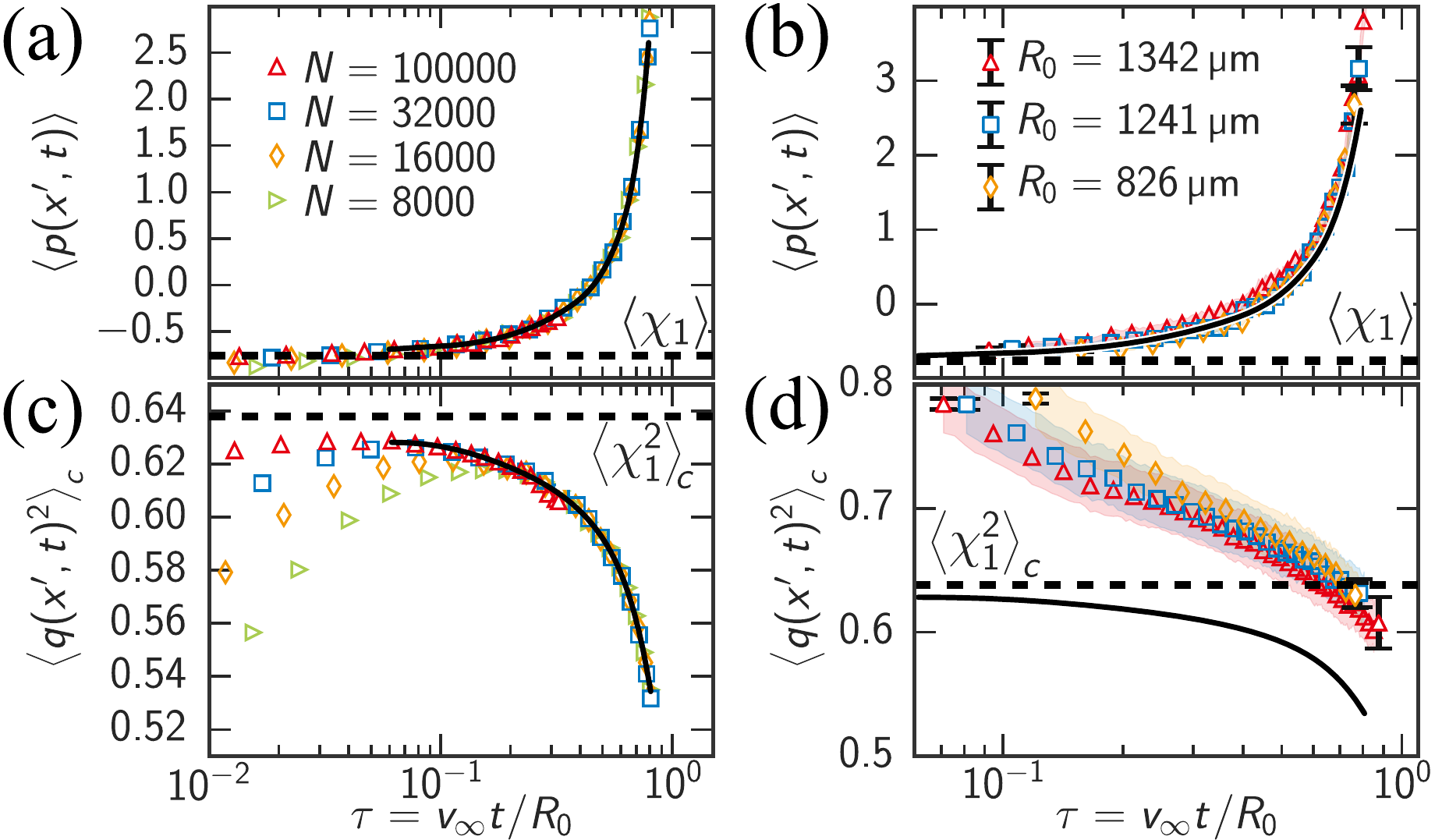}
\caption{\label{fig:time_rescaled}(color online). 
The mean rescaled velocity $\left<p(x,t)\right>$ (a), (b) and the variance $\left<q(x,t)^2\right>_c$ (c), (d) against rescaled time $\tau=v_\infty{t/R_0}$ for the Eden numerics (a,c) and for the LC experiment (b,d).
For the experimental data, the errors are indicated in the same way as Fig.\ref{fig:rescaled}(a,c). The black solid lines are the spline-curve fit of the Eden data with $t\ge10^3$.
The dashed lines show the mean or variance of $\chi_1$.
}
\end{figure}

The results so far indicate, both experimentally and numerically,
 that it is the \textit{flat} subclass that characterizes
 the ingrowing interfaces, until the deviation eventually occurs
 at some characteristic time.
Then what controls this deviation regime?
Let us remark here that ingrowing interfaces eventually collapse
 near the center of the ring,
 when $\left<h\right>\simeq{v_\infty}t$ reaches $R_0$.
This led us to rescale time as $\tau:=t/t_c=v_\infty t/R_0$
 with $t_c:=R_0/v_\infty$ being the (approximate) collapse time.
Figures \ref{fig:time_rescaled}(a) and \ref{fig:time_rescaled}(c) show the numerical data
 of the mean rescaled velocity $\left<p\right>$ and variance $\left<q^2\right>_c$,
 plotted against this rescaled time,
 and indeed the data collapsed well in the deviation regime.
It is reasonable that the deviation occurs near $t\approx{t_c}$,
 because the correlation length $\xi(t)\sim t^{2/3}$ then reaches
 the effective system size, or the circumference,
 $L(t)=2\pi(R_0-\left<h\right>)\simeq{2\pi\left({R_0-v_\infty{t}}\right)}$.
In contrast, in the short-time regime $t\ll{t_c}$, we expect that
 the asymptotic curves of the cumulants $\left<q^n\right>_c$,
 obtained in the limit $R_0, t\to{\infty}$ with fixed $\tau\ll{1}$,
 converge to $\left<\chi_1^n\right>_c$ as $\tau\to{0}$,
 which corresponds to the flat limit ($R_0\to\infty$).

We also made the same rescaling with the experimental data
 and found that $\left<p\right>$ against $\tau$
 overlaps on the curve of the Eden model within the error
 [Fig.~\ref{fig:time_rescaled}(b)],
 suggesting universality of this scaling function.
Concerning $\left<q^2\right>_c$, the experimental data do not overlap
 with the Eden curve [Fig.~\ref{fig:time_rescaled}(d)],
 but the deviation is smaller for larger $R_0$;
 therefore the two curves possibly overlap in $R_0\to\infty$,
 though the experimentally reachable value of $R_0$ is far too small
 to make a direct verification.

%%%%%%%%Summary%%%%%%%%

In summary, %combining the LC electroconvection and holography,
 we developed an experimental technique that allowed us to design
 the initial region of the DSM2 turbulence arbitrarily.
This opens a way to study KPZ interfaces from general initial conditions,
 experimentally, providing a basis for investigating
 the intriguing geometry dependence of the KPZ class.

Then we studied interfaces growing inward from a ring,
 both experimentally and numerically,
 using for the latter the off-lattice Eden model.
In the short-time regime ($t\ll{t_c}=R_0 /v_\infty$),
 our cumulant data %on the cumulants 
 [Figs.\ref{fig:skew_kurt} and \ref{fig:rescaled}(a)-\ref{fig:rescaled}(d)]
 agreed with those for the flat interfaces, 
 indicating that they are asymptotic to those of the GOE Tracy-Widom distribution.
In this regime the spatial correlation function was also approaching the $\text{Airy}_1$ correlation, which is another hallmark of the flat subclass. The circular subclass scenario was clearly ruled out.
Therefore, not only the presence of the curvature
 but its sign have a crucial effect
 on the determination of the universality subclass.
By contrast, in the long-time regime close to the collapse time
 $\approx{t_c}={R_0/v_\infty}$,
 the fluctuations were no longer controlled by the flat KPZ subclass.
The cumulants in this regime were found to be parametrized
 by dimensionless time $\tau=t/t_c$ (Fig.\ref{fig:time_rescaled}),
 implying the existence of scaling functions for this regime.
This long-time behavior is argued to occur
 as the correlation length $\xi(t)$
 becomes comparable to the effective system size $L(t)$.

Our results may arouse a few interesting discussions.
First, why does the sign of the initial curvature matter?
Although firm theoretical understanding needs to be made,
 our arguments based on the effective system size may give
 a tentative answer:
 while for the usual outgrowing circular interfaces
 the correlation length $\xi(t)\sim{t^{2/3}}$
 never attains the effective system size $L(t)\sim{t}$,
  in our ingrowing case, this does happen, as $L(t)$ now decreases in time.
Earlier numerical work \cite{i._s._s._carrasco_interface_2014}
 indeed showed that the system size expansion is crucial
 for the circular subclass.
Second, we note that ingrowing interfaces arise naturally in some systems,
 in particular in the coffee-ring experiment \cite{yunker_CLE_2013},
 where particles in a droplet accumulate onto the droplet edge.
This experiment needs to be revisited, because
 the data were compared with the GUE Tracy-Widom distribution.
There is also a model on the quantum entanglement entropy
 whose connection to KPZ was shown \cite{nahum_quantum_2017};
 in this model, the size of the growth region decreases in time,
 similarly to the ingrowing geometry we study here.
From broader perspectives,
 we remark that the split to universality subclasses occurs not only in KPZ,
 but also in another class of nonlinear growth problems
 \cite{carrasco_universality_2016-1}.
Analogous geometry dependence was also suggested
 for absorbing-state phase transitions \cite{lavrentovich_radial_2013}.
We hope our work will trigger further experimental and theoretical studies
 to elucidate this geometry dependence,
 which may be a new aspect of such out-of-equilibrium scaling laws.

\begin{acknowledgments}
We acknowledge discussions on the experimental and numerical methods with M. Sano, on theoretical aspects with T. Sasamoto, H. Spohn, P. Le Doussal, H. Chat{\'e}, on interpretations of the results with R. A. L. Almeida. We thank H. Spohn for having drawn our attention to Ref.\cite{nahum_quantum_2017}, suggesting possible connection to our work. We also thank F. Bornemann for the theoretical curve of the $\text{Airy}_1$ and $\text{Airy}_2$ correlation functions \cite{bornemann_numerical_2010} and M. Kuroda for technical support. Japan Society for the Promotion of Science (Grants No. JP25103004, No. JP16H04033, No. JP16K13846) and the National Science Foundation under Grant No. NSF PHY11-25915.
\end{acknowledgments}

\bibliography{citations}

\end{document}